# Winning Isn't Everything
## A contextual analysis of hockey face-offs

Nick Czuzoj-Shulman, David Yu, Christopher Boucher,

Luke Bornn, Mehrsan Javan

SPORTLOGiQ, Montreal, QC
nick@sportlogiq.com

## Abstract

This paper takes a different approach to evaluating face-offs and instead of looking at win percentages, the de facto measure of successful face-off takers for decades, focuses on the game events following the face-off and how directionality, clean wins, and player handedness play a significant role in creating value. This will demonstrate how not all face-off wins are made equal: some players consistently create post-face-off value through clean wins and by directing the puck to high-value areas of the ice. As a result, we propose an expected events face-off model as well as a wins above expected model that take into account the value added on a face-off by targeting the puck to specific areas on the ice in various contexts, as well as the impact this has on subsequent game events.

## 1. Introduction

It was October 28[th], 2017 and the Los Angeles Kings were in Boston, playing the 5[th] game of an early season 6-game road trip. The score was tied at 1 and with 0.9 seconds left in overtime, the face-off was in the Bruins' zone to the left of Tuukka Rask. The Kings pulled their goalie and sent Anze Kopitar out to take the draw against the Bruins' David Pastrnak. What happened next was truly incredible. With two right-handed shots lined up behind Kopitar, Drew Doughty behind him and Tyler Toffoli back and to the inside of the circle, the Kings center won the face-off clean and directly to Toffoli who immediately one-timed a shot past Rask and clinched the game with time still remaining on the clock. While it was a remarkable ending, there were some key elements that went into setting up the goal. For instance, heading into that face-off, Kopitar had won only 8 of 23 on the night, leaving him at an unimpressive 34.8% success rate. In comparison, Pastrnak, stuck out there after an icing call, had won his only other face-off of the night (100% success rate). Win percentages aside, Kopitar looked over his shoulder to see where his teammates were positioned and then cleanly directed the puck exactly where it needed to be for the Kings to get a scoring chance and ultimately win the game.

Going back a step, face-offs make up over 75% of all contested puck battles and despite being highly valued by most professional players, coaches, and management, face-off win percentages are not well correlated to goals or wins[1,2]. In fact, in the 2017-18 NHL regular season, they were -1.8% and 0.03% correlated, respectively. Additionally, it takes on average ~75 face-off wins to achieve a +1 goal differential[3]. Despite these underwhelming numbers, whenever it comes to a face-off, especially one like the high-pressured scenario described, the prevailing thinking tends to remain heavily focused on *win percentages*.

Looking at the example above, Kopitar was sent out on his weak side, when he was having a rough night in the circles, and still delivered significant value to his team. The Kings could have gone with Adrian Kempe, also a left-handed shot, who was 8-for-13 on the night (61.5%), or even Brooks Laich (9-for-16, 56.3%), but they went with their star player and as stars often do, he came through. The direction of the face-off win, the timing of it being clean, along with Toffoli's subsequent shot from a high danger area, and a bit of luck with Rask's reaction time made all the difference that night for an extra point in the standings.

## 2. Methods
### 2.1 Data Setup

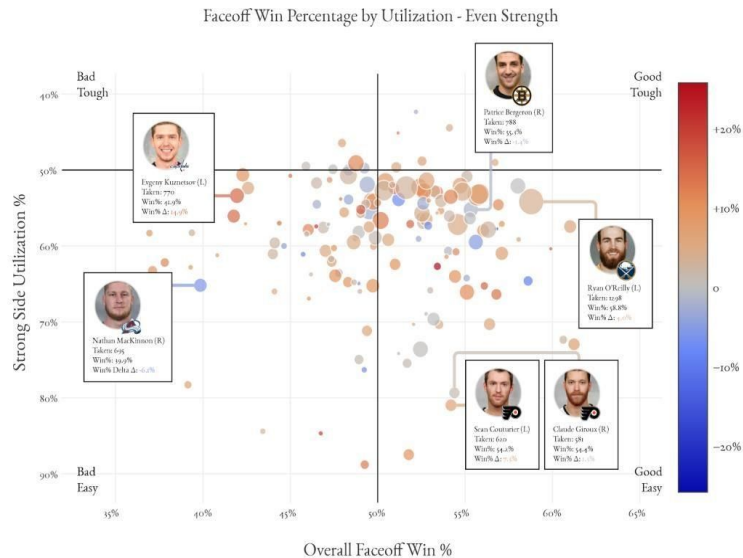

Figure 1 - Face-off win percentage by utilization in the 2017-18 regular season (minimum 200 face-offs). Marker size is proportional to number of face-offs taken. Color scale indicates the strong-weak difference in win percentages. While the majority of players win more face-offs on their strong side, there are some players (e.g. Nathan MacKinnon) who perform noticeably worse on their strong sides

We examine 71,147 face-offs and subsequent loose puck recoveries or pass receptions from the 2017-18 NHL regular season, capturing the spatio-temporal aspects of the face-off, as well as events such as shots (on net or from the slot) and zone exits/entries in the subsequent play sequences. The data comes from a proprietary SPORTLOGiQ set created via advanced computer vision techniques. A face-off is considered to be won or lost based on which team earns possession first after puck drop, and face-off wins can be marked as *clean* when the center wins the puck directly to a teammate, or *not clean* where a puck battle ensues or a player must skate from his stationary position to recover the loose puck that was not aimed directly toward him. Distance and time from puck drop to recovery/reception were measured along with player handedness and deployment, given the increasing trend toward utilizing players on their stronger sides (Figure 1). Strong and weak side face-offs were determined such that if the player's handedness matches the side of the boards he is closest to, then he is considered to be on his strong side. .

We develop metrics to measure face-off takers' ability to target the direction of the puck and show how such directionality leads to increased probability of generating a shot event (shot on net or attempt from the slot) or zone change. For each face-off circle, the puck directions were binned into 45° sectors so that a player can win the puck forward, backward, inside (toward the middle of the ice), and outside (toward the close-side boards), along with forward-inside, forward-outside, and backward-inside, backward-outside depending on the side of the ice the face-off takes place. Face-off locations were evaluated at each of the 9 possible circles on the ice, but for directionality and handedness results, the 12,285 (17.3%) center ice face-offs were excluded as they take place the furthest from high-value areas and could not be grouped with the rest of the circles based on inside/outside directions or strong and weak side deployments. Lost face-off zones and directions were reversed from the winner's side in order to properly represent the perspective of the losing draw-taker. For example, if a face-off took place in Team A's defensive zone, to the left of their goalie, and was won directly backward, then Team B's center would receive a lost face-off in the offensive zone, on the right side, that went in a forward direction. For zone analyses, the 3 zones were split into 4 sections: offensive zone, defensive zone, neutral zone on the attacking side of the red line (north), and neutral zone on the defending side of the red line (south). The neutral zone was split in this way to properly take team strategies into account and not bias subsequent zone entry numbers at the two different distances from the blue line. Center-ice draws were not considered in these zone section cases.

## 2.2 Model Methodology

Various approaches were taken in order to produce an Expected Events Adjusted Face-off Win Model (EE) and Win and Direction-Based Events Over Expectations (*WDBE*) metrics. The models look to subsequent play sequences for events such as zone entries/exits, shots on net or shots from the slot area of the ice and breaks them down by the direction that the face-off was won. All shot attempts from the slot were considered as *events* as this is the highest danger area of the ice, with a shooting percentage of 16.6% for the 2017-18 season compared to 9.0% for all shot locations. The Expected Events model was created by calculating a league-wide events average per face-off win direction, and multiplying this by the number of wins a player had in this same direction (*i* below) over the total number of face-offs he took.

$$EE = \sum_i \left[ \left[ \frac{\sum(Event_i)}{\sum(Faceoffs_i)} \right] * \left[ \frac{\sum(PlayerWinsTo_i)}{\sum(PlayerFaceoffs)} \right] \right] \qquad (1)$$

This approach, by using league averages, isolates the effects of the individual draw-taker from his teammates in order to better assess the value he creates by directing face-off wins to particular locations.

The Win-Based Events Over Expectation Model considers the league average events per face-off win, and per overall face-off (win or loss), whether the player was on his strong side and if the win was clean, and then breaks them down into the 4 zone groups outlined in Section 2.1. Next, we subtract these deployment and win-type adjusted event averages per win by the averages for face-offs won or lost of the same category and multiply this by 2 in order to produce an expected number of events for a win vs a loss. This number is finally then multiplied by the player's individual probability of winning the face-off in each direction ($i$), given his deployment ($s$) and win-type ($c$), which is also broken down by zone section ($j$ below).

$$WDBE = \sum[P(WinDirection_{jisc}) * E(Event_{jisc})] \qquad (2)$$

The EE model represents an aggregated overview of a face-off taker's ability to target important areas off a face-off, without penalizing him for the quality of his teammates, and without separating his performance by zone or side of the ice. In comparison, the WDBE model incorporates the player's expected events but goes further in order to capture the added benefits of clean wins and differing abilities/strategies depending on the location of the face-off, based on both the zone and the side of the ice.

## 3. Results
### 3.1 Clean Face-offs

#### 3.1.1 Clean Face-offs Breakdown

Hockey is a game of limited time and space and due to the high-end talent that many of the players possess, finding ways to create additional time and space for your team will typically lead to more positive outcomes. As outlined in the Kopitar play in Section 1, a clean win is one such way of doing this. In our data set, 44.5% of face-off wins were counted as clean. Of those clean wins, 48.0% led to a shot event or zone exit/entry compared with 44.1% from non-clean wins. While these numbers don't differ greatly, some key differences do emerge when digging into different zones. In the offensive zone, for example, clean wins led to a shot event 38.6% of the time compared to 30.3% for non-clean ones (Figure 2). Taking this further and looking at directions, clean wins that were directed backward-inside (such as the Kopitar to Toffoli play) led to shot events 43.6% of the time vs. 32.1% off a non-clean win. Of note for the 2017-18 NHL regular season: clean wins, on average, were directed to a teammate's stick 25% faster than non-clean wins. When a player wins the draw cleanly, his team can execute drawn-up set plays with greater ease and will therefore have better chances of catching the defending team off-guard or out of position.

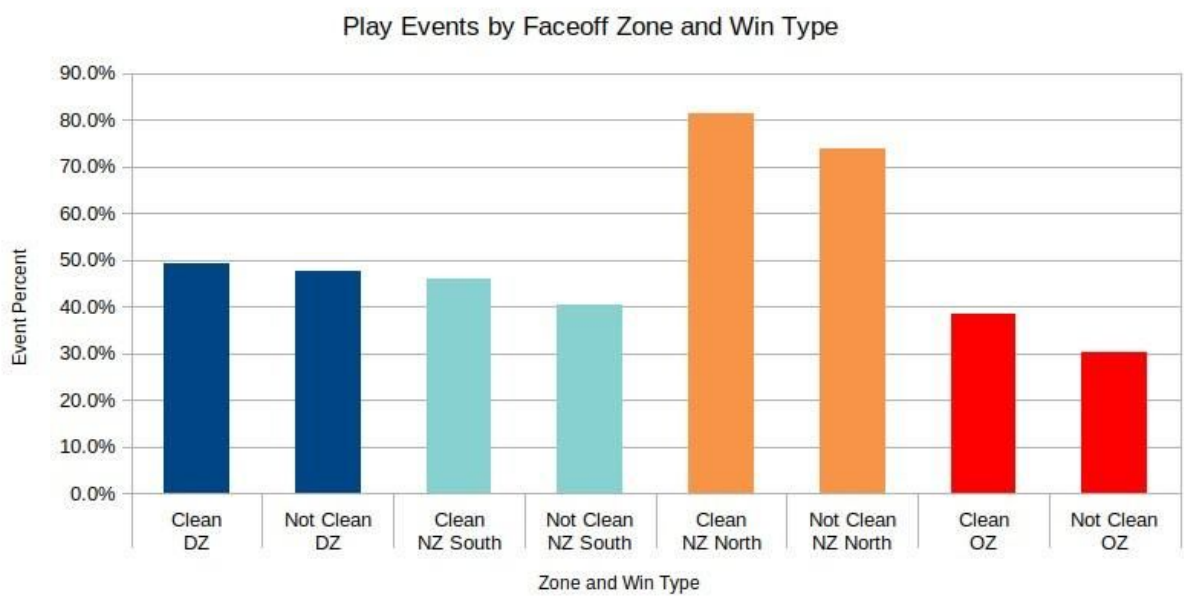

Figure 2 - Percentage of events after face-off wins by zone and win-type. Clean wins have higher percentages of events in every zone, however the gaps are largest on the attacking side of the ice. The smallest difference is in the defensive zone.

Looking at defensive zone face-offs split by clean and non-clean wins tells an interesting story about the prevailing strategy when facing off inside ones' blue line. In the defensive zone, wins with zone exits occur 49.2% of the time off a clean win, compared to 47.5% for a non-clean one (Figure 2). The rationale behind this similarity in numbers is that teams are aware of the importance of winning a face-off clean in the offensive zone and will have their center approach the face-off in a way where he limits his chances of losing it clean. As a result, more defensive wins will be from tie-ups and other plays such as winger or defensive support. In fact, only 35.4% of defensive zone wins were done cleanly in our sample compared to 50.4% of wins that were clean in the offensive zone (Figure 3). Neutral zone wins also come close to 50% clean and so the key here is the effort made by defending centers.

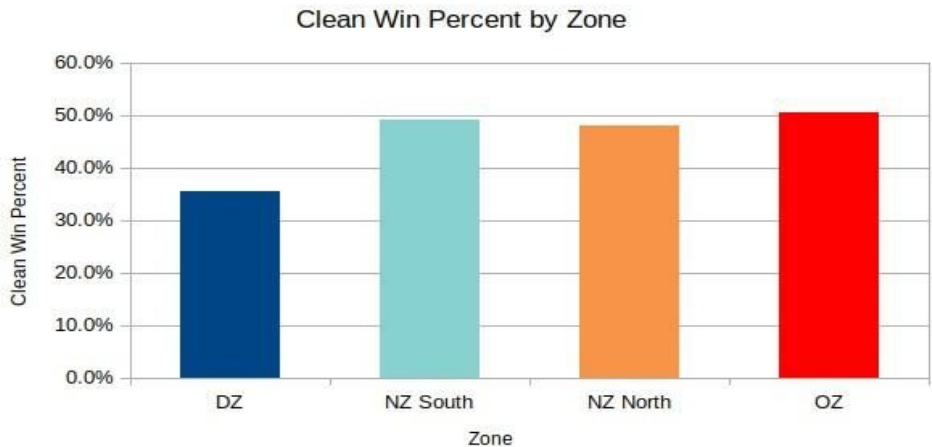

Figure 3 - Percentage of wins that are clean broken down by zone. Offensive zone has the highest percentage of wins that are clean, whereas in the defensive zone, more wins come from a non-clean approach as teams try to limit the clean wins against them. Neutral zone on the attacking or defending side of the red line are similar to the offensive zone.

### 3.1.2 Player Level Clean Face-offs

At the player level, some players demonstrate a high level of ability at winning the draw clean, whereas others display clear tendencies toward relying on tie-ups, help, and other non-clean methods of winning possession. Among players with 200 or more face-offs taken, Brian Boyle led the league for the 2017-18 season with 57.3% (Table 1) of his wins as clean ones, while Mark Jankowski was at the other end of the list with 31.6% clean wins (Appendix 3.1). There is a high likelihood that team strategies played a part in the lower outcomes here as 3 of the bottom 4 centers in clean win percentage played for the Calgary Flames, but at the top end it does appear to be more talent-based. The Flames also went out this past the off-season and acquired Elias Lindholm, who ranked in the top 20 in this category last year. Lindholm's winning percentage among clean face-offs also ranked him in the top 4, meaning that when a face-off result was clean, he was on the winning end 60.1% of the time.

Looking at the above table it is again clear that overall face-off win percentages do not tell enough of the story when assessing a player's skills in the circles. In fact, the ability to win the draw clean among all a player's wins has a fairly negligible correlation with their overall face-off winning percentage, at 2.0%. In the subset of our data among centers who took 200 or more face-offs, the overall face-off win percentages had a -7% correlation with Expected Goals Differential WOI (XGD), a significant measure of quality shot attempts for and against while a player is on the ice. This measure takes into account the location and context surrounding a shot for or against, in order to assign the differential in goal expectancy while a player is on the surface. In contrast to overall face-off success, the percentage of a player's wins that were clean correlated with XGD at 23.8% thus supporting the importance of clean wins and the players that can achieve them more often. Whether there are 0.9 seconds left on the game clock or more, winning a face-off cleanly has a meaningful impact on the ensuing events on the ice.

| Player | Clean % Among Wins | Win % Among Clean | Overall Win % |
|---|---|---|---|
| Brian Boyle | 57.3 | 50.8 | 51.5 |
| David Perron | 54.8 | 49.5 | 41.7 |
| Claude Giroux | 54.8 | 60.3 | 53.8 |
| Evander Kane | 54.5 | 50.8 | 43.7 |
| Dylan Strome | 54.4 | 45.0 | 45.0 |
| Henrik Zetterberg | 53.8 | 49.3 | 48.3 |
| Ryan Johansen | 53.5 | 58.6 | 55.8 |
| Gabriel Landeskog | 52.4 | 55.6 | 47.5 |
| Anze Kopitar | 52.3 | 58.3 | 54.9 |
| David Krejci | 52.3 | 56.0 | 51.5 |
| Nolan Patrick | 52.3 | 54.9 | 50.4 |
| Anthony Cirelli | 52.2 | 45.6 | 44.3 |
| Pierre-Edouard Bellemare | 51.6 | 53.8 | 53.0 |
| Jonathan Marchessault | 51.5 | 46.3 | 42.9 |
| Vincent Trocheck | 51.1 | 55.8 | 55.1 |
| Andrew Shaw | 50.9 | 61.7 | 55.7 |
| Elias Lindholm | 50.8 | 60.1 | 56.8 |
| Jordan Staal | 50.8 | 59.0 | 55.4 |
| Dylan Larkin | 50.8 | 51.2 | 49.0 |
| Kyle Turris | 50.7 | 52.4 | 49.6 |

Table 1 - Top 20 players by percentage of face-off wins that were clean for the 2017-18 NHL season, minimum 200 face-offs taken. The centers at the top of the list display a clear skill in winning face-offs cleanly, adding value while they are on the ice. The overall face-off win percentage is included to show that players like David Perron, who would be thought of as inferior face-off takers, rank highly in more significant measures of face-offs. Win percentage among clean face-offs shows the differential between wins and losses when the face-off outcome is clean.

### 3.2 Face-off Directionality

### 3.2.1 Directionality Breakdown

The LA face-off example above also displayed how value can be created on the ice by directing the puck to a specific target and allows us to show how some players have the ability to do this consistently. At the 8 face-off circles (excluding center-ice), face-off wins were pushed backward most often, nearly ⅓ of the time (Figure 4). Interestingly, wins were also directed more in the backward-outside (toward the boards) direction than to the backward-inside (to the center of the ice), and the same was true for straight outside as compared to straight inside. One explanation here is that directing the puck to the outside will likely maximize the time your team has to begin a play. By examining average elapsed time from puck drop to a face-off win reception, this idea holds true. In fact, backward-outside face-off wins were more than 16% faster when compared to backward-inside ones. Additionally, teams will strategize to keep each other to the outside of the ice as much as possible, away from the more dangerous areas, and so winning a face-off back or back and to the outside will therefore mean a maximum amount of time for your team to set up and organize as defenders will approach from the inside out. Deployment and strong-handedness also apply to the timing and direction preferences here as a strong side deployed player drawing the puck back and outside will very likely be using a quick backhand swing to accomplish this. On the other hand, a backward-inside win on a player's strong side of the ice would involve a greater (and slower) body rotation to push the puck back in this direction. There is more on deployments and player handedness in Section 3.3.

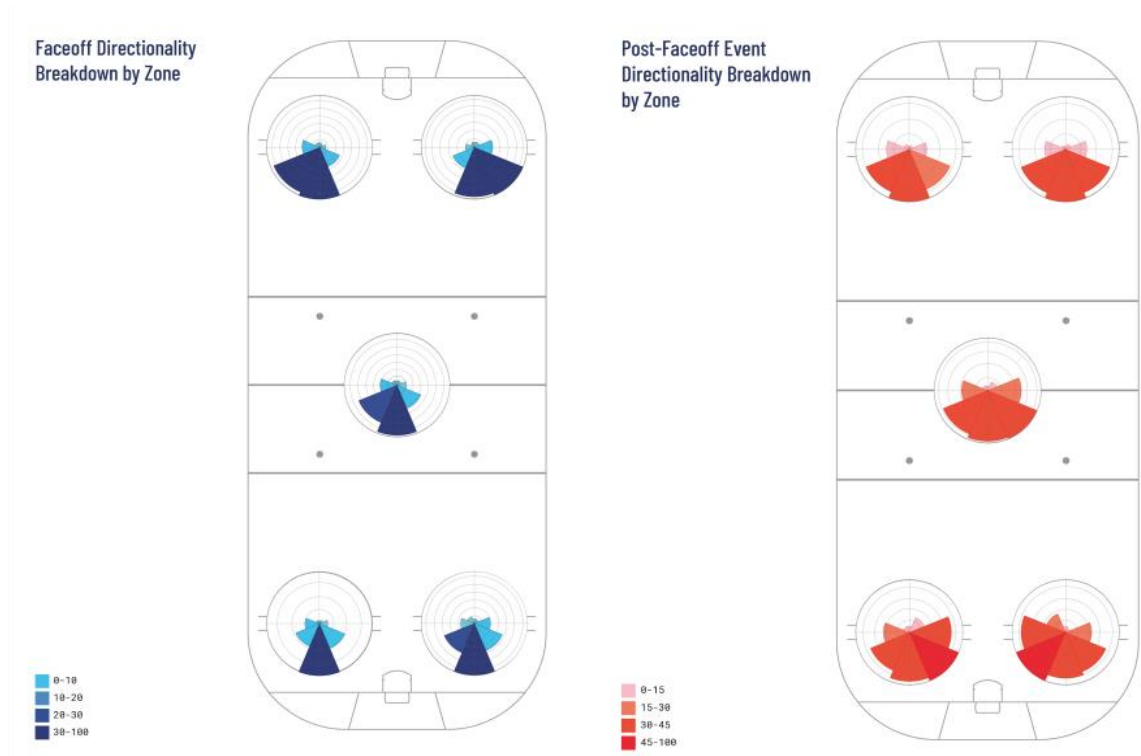

Figure 4 - On the left: Proportion of face-off win directions by location on the ice. The center of both charts on the left and right are not center-ice, but rather the overall proportions for all zones combined. Center-ice draws are excluded from this analysis as there are no true *inside* or *outside* directions here. In the offensive zone, higher proportions of wins are directed backward or backward-outside, whereas in the defensive zone, the backward angled wins are more even and straight-backward remains highest. Not shown here are neutral zone draws where trends similar to the offensive zone appear. On the right: The distribution of post-face-off shot and zone change events by zone and directionality is shown where backward and backward angled targets lead to more events. Petal size and colour correspond to proportions of the face-offs at each location that are won in each direction (left) or lead to game events (right).

If we break the win directions down by their individual locations on the ice, more trends emerge. For instance, on the left side in the offensive zone, the proportion of wins directed backward and backward-outside become nearly identical, at 33.1% and 31.3% respectively. The same holds true on the right side in the offensive zone with 30.4% backward and 31.4% wins pushed backward-outside. Wins to the inside or forward are also much less common as these directions have a significantly higher chance of leading to turnovers. Simply based on how teams line up, these forward and sideways wins generally mean that a second puck battle will take place, given that each attacking player will have a defender much closer to him as compared to backward directions where defenders are forced to line up further away. In the defensive zone, winning the puck backwards again occurs most frequently, however in this zone, it's the backward-inside and backward-outside directions that occur nearly the same. On the right side in the defending zone, face-offs were won 20.2% backward-inside and 18.7% backward-outside. Whereas on the left, these numbers were 19.8% and 18.1% respectively. The explanations for this are likely also affected by player handedness and team strategies, but overall defending teams appear to strategize to hit the puck to the corner or behind the net in order to set up and start their breakouts.

### 3.2.2 Directionality and Subsequent Events

In looking at subsequent events after a face-off, directionality plays a major role. Of all events that occurred after a face-off win, more than ⅓ of these came from backward drawn pucks. Given that this is the direction that ⅓ of all face-offs are won, we opted to look among the wins targeted in each direction to see how many were resulting in shot or zone change events. Observing the within-direction event percentages, it's clear that not all targeted areas from a face-off win have the same values (Figure 4). Directing the puck backward had a 41.3% chance of leading to an event, whereas backward-inside and backward-outside wins were not far behind at 42.7% and 39.0% respectively. Inside and outside targets were a little more than half of this, while moving the puck forward off a face-off proved again to be significantly worse, with generated shot and entry/exit plays occurring less than 10% of the time.

Breaking these numbers down by zone, backward and backward angled pucks had more than double the chances of leading to shot events for teams in the offensive zone, as compared to sending the puck laterally inside/outside. They also had nearly 10 times the chances of events compared to aiming the puck in one of the forward positions. Taking this further and comparing clean to non-clean wins in the offensive zone is even more telling. A non-clean face-off won back and to the inside led to a shot event close to 19% of the time, whereas a clean draw (like the Kopitar play above) more than doubles this likelihood and had a subsequent shot event more than 40% of the time.

In the defensive zone, a similar trend existed with sending the puck backward and to the inside leading to the most zone exits. The reasoning behind this is that defensemen recovering or receiving the puck off these face-offs were able to protect themselves behind the net, leading to additional time for their teammates to organize and for them to make a less rushed first pass when starting the breakout. One surprise here, although likely due to a variety of factors, was that winning the puck to the inside in the defensive zone resulted in a relatively high number of exits. It's important to note that these were inside draws where the team had successfully gained possession and so assuming the players on the defensive side were alert for these face-offs, the risk of a target here could be quickly remedied by skating or passing the puck to a teammate outside of this otherwise high danger area. A quick knock back to the outside defenseman making his way behind the net is just one example of such a remedy. Taking this one step further, inside wins that led to zone exits occured 26% slower on average than inside wins with no exit. The likely explanation here is that the defensive center tied up his opponent, knowing that his inside winger was coming in to support and start the exit, thus taking additional time from puck drop.

In the neutral zone, on the defending side of the red line, the backward directions continued to be the best options. Wins to the inside were not as good as they were in the defensive zone, and pushing the puck forward was still not recommended. On the attacking side of the neutral zone, the direction results were the same, but the numbers grew significantly. Backward directed wins produced zone entries more than 60% of the time, and the number of inside/outside targeted entries were also much higher at 45.5% and 37.4% respectively.

### 3.2.3 Player Level Directionality

| Player | Win Direction | Expected Event % | CF % | XGD |
|---|---|---|---|---|
| Jay Beagle | Backward | 12.3% | 31.8% | 28.9% |
| Luke Glendening | Backward | 11.0% | 37.5% | 36.1% |
| Antoine Vermette | Backward | 10.4% | 46.5% | 46.7% |
| Colton Sissons | Backward | 10.0% | 48.5% | 46.6% |
| Brandon Sutter | Backward | 9.9% | 38.7% | 34.8% |
| Andrew Shaw | Backward | 9.7% | 56.5% | 57.9% |
| Ryan Johansen | Backward | 9.1% | 56.9% | 55.1% |
| Brandon Dubinsky | Backward | 9.1% | 43.9% | 41.9% |
| Lars Eller | Backward | 8.7% | 46.3% | 44.1% |
| Claude Giroux | Backward | 8.7% | 60.2% | 63.9% |
| Victor Rask | Backward | 8.7% | 57.6% | 59.7% |
| Nick Bonino | Backward | 8.6% | 43.7% | 42.4% |
| Ryan O'Reilly | Backward | 8.6% | 54.9% | 54.8% |
| Adam Lowry | Backward | 8.5% | 53.0% | 54.1% |
| Joe Pavelski | Backward-Outside | 8.5% | 59.1% | 61.4% |
| Boone Jenner | Backward | 8.5% | 46.2% | 47.4% |
| Vincent Trocheck | Backward | 8.5% | 53.1% | 51.8% |
| Sean Couturier | Backward | 8.5% | 57.0% | 58.9% |
| Marcus Kruger | Backward | 8.5% | 48.5% | 40.7% |
| Kyle Brodziak | Backward | 8.4% | 43.3% | 39.5% |

Table 2 - Player expected events per face-off win direction, along with their Corsi For percent and Expected Goals While On Ice Differential. Players who won the majority of their face-offs backward had the highest expected events. Win direction here was the targeted angle off a face-off win that led to the highest expected events for the player.

At the player level, an expected events model was produced (Section 2.2, equation 1) to tie all the above together in evaluating a player's true face-off ability and resulting impact on a game. This model used league average shots on net or attempts from the slot, along with defensive zone exits and offensive zone entries in the plays following a face-off, broken down by direction of a draw win, as weights for the number of wins a player had in each of these directions. Unsurprisingly, players who directed their wins backwards or backwards-outside most often came out on top of the list (Table 2). Among the subset of centers taking 200 or more face-offs, the EE metric correlated with Corsi For percentage (CF%) at 6.6% and with XGD at 7.3%. CF% is a way of measuring all shot attempts for and against while a player is on the ice, approximating which way the player in question may be driving the play while he's on the ice. XGD assigns value and quality to shots as per section 3.1.2.

Some players showed a unique ability to direct pucks to either side of the ice, regardless of where the face-off took place (Figure 5). Sidney Crosby, a left-handed shot, is one example who did just that. Comparing Crosby's face-off win directionalities against those of Riley Nash, a right-handed shot, we can see how Nash preferred to win the majority of his draws back and to the right, likely using the backhand of his stick.

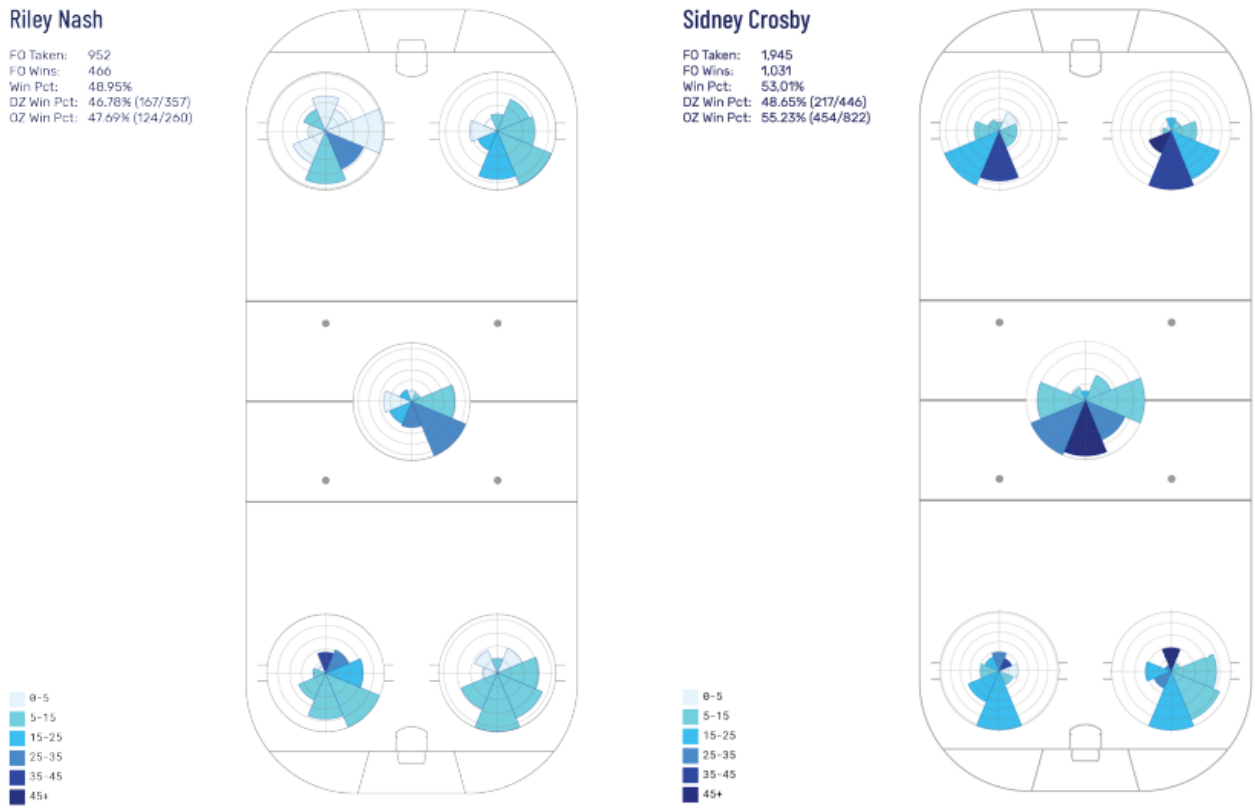

Figure 5 - Face-off win directionality by circle. Petal size is proportional to the number of face-off wins, while colour indicates the average distance in the given direction, per face-off circle and player. Directions were binned in 45-degree sectors and represent a clear indication of where the player/team show preferences in strategy/skill at each of the face-off circles. We see that in the OZ, Nash consistently pulls back and to the right, while Crosby is able to pull to the outside point from both OZ circles.

### 3.3 Face-off Deployment

#### 3.3.1 Directionality and Deployment

In our sample of NHL face-offs, strong side deployments were significantly more frequent (61.0%) for the 2017-18 season. This is very likely due to more teams recognizing the improved success, on average, their players have when deployed favourably. In fact, centers on their strong sides won 11.1% more often than when taking draws on their weak sides. In looking at which directions were targeted most often by player deployment, the mechanics of a face-off really stood out (Figure 6). Players on their strong sides typically were able to take advantage of the strength of their lower hand as well as the full blade of their stick, which is more natural to do when attempting to propel the puck backwards or backwards-outside. This is outlined by the fact that roughly 65% of strong side face-offs were won in these two directions combined. By comparison, the same two areas were targeted less than 50% of the time on weak side deployments. Players on their weak sides tended to favour their backward-insides more often than the outside, meaning that they still preferred using the backhand of the stick here even if it meant directing the puck to the middle of the ice. These trends held similar when the analysis was split up by zones (Appendix 3.3).

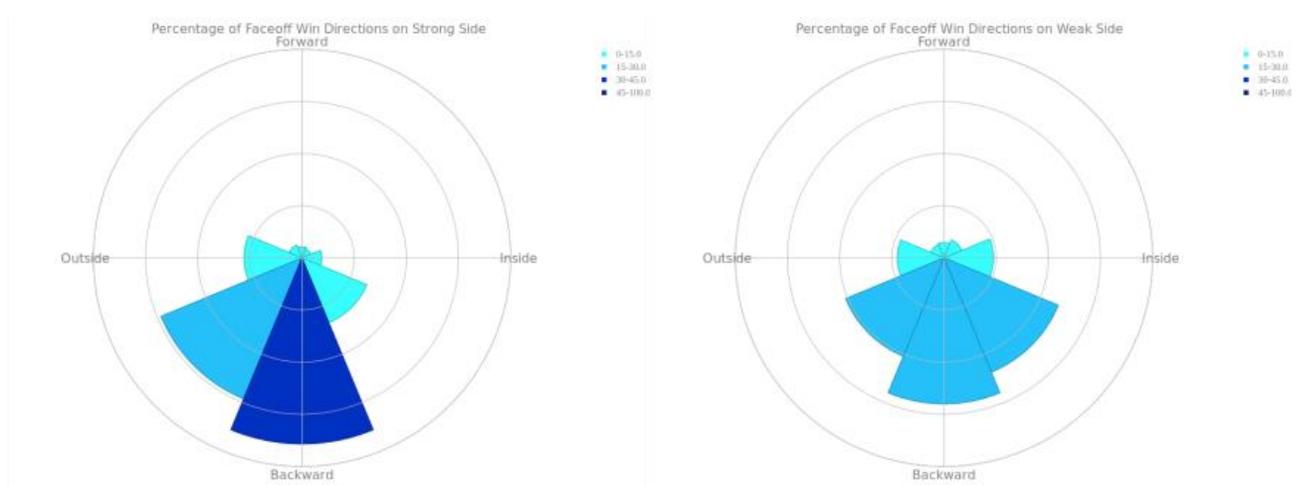

Figure 6 - Percentage of face-off win directions broken down by strong and weak side deployments. Strong side players favoured the backward and backward-outside areas, utilizing the more natural backhand movements of their stick and strength from their lower hands. Weak side deployed centers still used their backhands, and thus showed a greater propensity for directing the puck backward-inside as compared to when on the strong side. Forward and lateral directed wins remained low.

#### 3.3.2 Deployment and Subsequent Events

Subsequent game events following a face-off remained fairly similar whether the face-off-taker was on his strong or weak side, with a few noticeable exceptions (Appendix 3.3.2). The similarities were likely due to the arguments laid out in Section 3.2.2 where directionality is thought to have a bigger impact on generating an event as compared to deployment factors. Given that centers can be thrown out of face-offs for violation purposes, and the fact that players will not always be facing off on their strong sides, they still need to find ways to direct the puck to valuable areas and overcome the lack of a favourable opportunity. That being said, the Kopitar scenario once again comes into play to outline where handedness can have a substantial role. The Kings center was on his weak side in the offensive zone for that face-off, and thus used his backhand to direct the puck backward-inside for a valuable shot on net from the slot. Not surprisingly, there was a trend of this where weak side deployed backward-inside wins in the offensive zone, which we have already seen occurring more often and leading to more events when won cleanly, had a 33.7% chance of leading to a shot event as compared to only 26.7% on the strong side. The same concept works for the attacking side of the neutral zone, where weak side backward-inside wins led to 7% more zone entries. The fact that players can win these draws to this valuable area of the ice on their backhand gives them greater flexibility even if unfavourably deployed and speaks to the creative strategies teams work on to generate value no matter the circumstances.

## 4. Aggregated Player Models

The above expected events model works on a relatively simple level to display who is winning their face-offs to higher value areas of the ice and while it does a better job than overall winning percentages, it still correlates weakly with valuable player metrics like XGD. Further to this, it doesn't adjust for the side of the ice where the face-off took place, which we have seen has a meaningful effect on where players generally direct their wins, and similarly doesn't adjust for the type of win, which we have seen is a powerful indicator for the likelihood of a subsequent event occurring. As a result, we propose the Win and Direction-Based Events Over Expectation Model (WDBE, equation 2) to account for these factors and also take the 4 zone sections of the ice into account. Early versions of this model, which only accounted for zone and directions, found an 8.0% correlation with XGD in the offensive zone and 6.0% for the defensive zone, but overall didn't improve greatly on the EE approach. This method was therefore replaced by the deployment and win-type adjusted version which resulted in more meaningful results (Table 3). In the offensive zone, adjusted WDBE's correlated with XGD at 13.1%, this association was 10.5% in the defensive zone, and 7.2% and 11.1% on the attacking and defending sides of the neutral zone respectively.

| Player | WDBE % | CF % | XGD | Clean % Among Wins | Win % Among Clean | Overall Win % | Strong-Side % |
|---|---|---|---|---|---|---|---|
| Joe Pavelski | 89.0% | 59.1% | 61.4% | 47.2% | 54.2% | 54.3% | 89.3% |
| William Nylander | 88.8% | 56.2% | 56.7% | 36.3% | 51.2% | 50.5% | 89.3% |
| Evander Kane | 78.2% | 55.7% | 56.2% | 54.5% | 50.8% | 43.7% | 89.7% |
| Claude Giroux | 76.1% | 60.2% | 63.9% | 54.8% | 60.3% | 53.8% | 75.3% |
| Jonathan Marchessault | 70.0% | 58.8% | 62.2% | 51.5% | 46.3% | 42.9% | 75.2% |
| Patrick Marleau | 60.7% | 53.3% | 54.0% | 42.9% | 52.7% | 55.3% | 78.7% |
| Gabriel Landeskog | 60.6% | 52.5% | 54.8% | 52.4% | 55.6% | 47.5% | 85.4% |
| Jamie Benn | 58.6% | 55.5% | 61.8% | 50.1% | 52.3% | 50.4% | 77.3% |
| Luke Glendening | 54.4% | 37.5% | 36.1% | 34.8% | 54.4% | 58.8% | 75.7% |
| Ryan Carpenter | 53.7% | 48.6% | 43.7% | 38.2% | 47.9% | 49.2% | 86.1% |
| Sean Couturier | 53.7% | 57.0% | 58.9% | 46.8% | 55.0% | 53.7% | 78.9% |
| Charlie Coyle | 49.3% | 46.2% | 52.6% | 43.0% | 50.5% | 50.5% | 78.5% |
| Auston Matthews | 47.4% | 55.3% | 58.4% | 33.9% | 52.3% | 54.6% | 77.0% |
| Nathan MacKinnon | 47.4% | 56.1% | 59.0% | 50.3% | 42.4% | 42.4% | 69.9% |
| Victor Rask | 46.1% | 57.6% | 59.7% | 44.7% | 56.1% | 56.1% | 73.1% |
| Chris Wagner | 45.2% | 39.1% | 34.6% | 39.1% | 44.2% | 49.3% | 72.1% |
| Elias Lindholm | 44.4% | 56.9% | 57.5% | 50.8% | 60.1% | 56.8% | 77.2% |
| Steven Stamkos | 43.5% | 58.8% | 62.6% | 41.9% | 49.5% | 50.9% | 74.9% |
| Tyler Seguin | 43.3% | 54.2% | 60.0% | 45.9% | 52.4% | 53.8% | 76.3% |
| Mark Letestu | 42.8% | 49.7% | 49.2% | 36.7% | 53.4% | 53.7% | 74.3% |

Table 3 - Top 20 players from the WDBE face-off model, adjusting for strong side usage and clean win percentages, per zone. This model found stronger relationships with individual player metrics and gave more weight to events after the face-off than relying on winning percentages alone. Players who were deployed heavily on their strong sides were able to generate significant value in their favourable opportunities.

At the other end of the list (Appendix 4), we can see players who failed to generate as much value for their teams, whether due to a lack of clean wins, an inability to direct pucks to high impact areas (regardless of deployment opportunities), or simply just not winning enough face-offs. Looking back at the top end of the WDBE group, we can see how San Jose utilized players such as Joe Pavelski and Evander Kane (despite the latter's lower overall win percentage) highly on their strong sides and were rewarded for doing so. There is a significant drop off in WDBE % suggesting that there's plenty of room for players to continue to improve upon their skills and work on generating more value, but it's also worth mentioning team strategies here. San Jose, as an example, has made a considerable effort to deploy their centers on their strong sides[5] and has been on the upper end of goal differentials after offensive zone draws[6]. These strategies likely played a role in Pavelski and Kane's overall scores.

## 5. Discussion

While the intentions and skill levels of centers in the face-off circles have been measured to new and highly granular levels, it is still worth noting some of the drawbacks that were encountered. One such example is that all face-off wins result from one or two sticks colliding with a puck that was dropped from above by a human with lots of possibility for errors and randomness. The drop could be slightly more to one side or the other, it could bounce strangely due to the shape of the puck and the movement of sticks below it, or it can ricochet off either centers' skates or the skates of the linesman involved. All this to say that some directions of face-off wins will be directly related to these random factors, but with a sample size of a full season worth of puck drops, the impact of these random outcomes should be minimized allowing for true player skills and intentions to be observed.

Another point worth noting is that teams may win immediate possession of a face-off but then lose control of the puck just as quickly. In this case the team and player would still be credited with a face-off win and the resulting timing/directionality results would be assigned along with no subsequent play event, but no additional factors would be considered. This is especially important for face-offs where the center wins it forward or laterally, into a dangerous area only to have his team lose possession and possibly surrender a scoring chance or play against immediately after. Similarly, plays where the center tries to win it in a certain direction, but has it intercepted by an opponent would result in a win for the opposing center in the direction that the puck was picked up.

Other, less measured, factors that could influence how players operate at the face-off dots include team strategies, which were touched on briefly, but also player predictability within a game. For example, if the same players go head-to-head repeatedly throughout a game, chances are they will try to mix up their techniques to avoid giving their opponents a pattern to follow and ultimately the upper hand. This could also result in different directions being targeted as a game wears on, or perhaps additional losses in areas deemed less valuable, as players try mixed or higher risk maneuvers.

## 6. Conclusion

While the value of individual face-offs is thought to be small, we find that there is considerable variability in value based on game context (e.g. the zone and side of the ice where the face-off takes place, along with player handedness and usage). We show that even within the same game context, not all face-off wins are created equal, with some players creating additional value through clean wins and targeting specific areas on the ice. In fact, players who generate the most value from face-offs do not necessarily have higher face-off win rates, suggesting that both win percentages and win value must be taken into account to measure the full impact generated during a draw. Clean wins were shown to be faster and considerably more valuable, especially on the attacking side of the ice, whereas strong and weak side deployments provided better indications of where the puck may be directed. Overall, individual players who can target pucks to the highest value areas on the ice, with a clean approach, were shown to produce the most value for their teams and ultimately redirect the way face-offs can be viewed. Our newly proposed metrics allow us to measure this post-face-off value for the first time, enabling us to understand the full value a player creates in the face-off circle beyond simple win percentages.

## 7. Acknowledgements

The authors would like to acknowledge Philippe Desaulniers, Evin Keane, Matt Perri, Sam Gregory, Michael Horton, and the Design and Eventing Ops Teams at SPORTLOGiQ for their efforts and contributions to this project.

# Appendix

| Player | Clean % Among Wins | Win % Among Clean | Overall Win % |
|---|---|---|---|
| Logan Couture | 37.2 | 45.1 | 45.5 |
| Nick Foligno | 37.1 | 43.9 | 48.9 |
| Mark Letestu | 36.7 | 53.4 | 53.7 |
| Cristoval Nieves | 36.7 | 46.8 | 49.0 |
| Pavel Zacha | 36.6 | 41.1 | 44.6 |
| William Nylander | 36.3 | 51.2 | 50.5 |
| David Kampf | 36.1 | 51.8 | 53.5 |
| Byron Froese | 35.5 | 44.4 | 48.3 |
| Tomas Hertl | 35.0 | 47.9 | 52.5 |
| Luke Glendening | 34.8 | 54.4 | 58.8 |
| Tommy Wingels | 34.7 | 36.8 | 45.7 |
| Jacob Josefson | 34.7 | 36.8 | 42.9 |
| Carl Soderberg | 34.6 | 39.8 | 46.4 |
| Kyle Brodziak | 34.6 | 47.7 | 53.6 |
| Andreas Athanasiou | 34.3 | 39.1 | 40.2 |
| Auston Matthews | 33.9 | 52.3 | 54.6 |
| Mikael Backlund | 33.7 | 48.4 | 53.7 |
| Matt Stajan | 33.5 | 41.9 | 49.3 |
| Tyler Bozak | 32.2 | 54.2 | 52.8 |
| Mark Jankowski | 31.6 | 43.2 | 48.3 |

3.1 - Bottom 20 players based on percentage of clean wins among face-offs won (minimum 200 face-offs taken). Team strategies are likely a factor here, given the number of players on the same teams appearing, however, the value in clean face-off wins should not be ignored.

| Deployment | Direction | DZ Win Dirction Breakdown % | NZ South Win Direction Breakdown % | NZ North Win Direction Breakdown % | OZ Win Dirction Breakdown % |
|---|---|---|---|---|---|
| Strong | Backward | 40.4 | 33.6 | 29.3 | 32.3 |
| Strong | Backward-Inside | 16.9 | 15.7 | 12.9 | 9.2 |
| Strong | Backward-Outside | 20.5 | 35.0 | 38.4 | 35.7 |
| Strong | Forward | 2.1 | 1.0 | 0.9 | 2.7 |
| Strong | Forward-Inside | 2.4 | 1.3 | 1.5 | 1.5 |
| Strong | Forward-Outside | 2.4 | 1.3 | 2.2 | 3.4 |
| Strong | Inside | 4.6 | 4.2 | 4.8 | 2.7 |
| Strong | Outside | 10.8 | 7.9 | 10.0 | 12.4 |
| Weak | Backward | 27.1 | 27.2 | 21.8 | 30.8 |
| Weak | Backward-Inside | 25.5 | 27.6 | 26.1 | 20.2 |
| Weak | Backward-Outside | 14.5 | 25.8 | 27.6 | 23.5 |
| Weak | Forward | 3.3 | 1.1 | 1.2 | 3.5 |
| Weak | Forward-Inside | 5.7 | 1.9 | 2.1 | 2.5 |
| Weak | Forward-Outside | 2.8 | 1.2 | 1.7 | 3.4 |
| Weak | Inside | 12.0 | 8.3 | 10.7 | 6.7 |
| Weak | Outside | 9.1 | 7.0 | 8.7 | 9.4 |

3.3 - Face-off win directions by deployment and zone. Above is a breakdown of the directions that face-offs are won, split by strong and weak side deployments. For the most part, these numbers don't differ too much, with a few exceptions such as backward-inside and backward-outside in the OZ or DZ.

| Direction | Weak Side DZ Event % | Strong Side DZ Event % | Weak Side NZ South Event % | Strong Side NZ South Event % | Weak Side NZ North Event % | Strong Side NZ North Event % | Weak Side OZ Event % | Strong Side OZ Event % |
|---|---|---|---|---|---|---|---|---|
| Backward | 41.0 | 42.6 | 44.6 | 41.5 | 69.6 | 71.3 | 34.0 | 34.8 |
| Backward-Inside | 47.1 | 45.3 | 44.7 | 37.4 | 75.0 | 69.6 | 33.7 | 26.7 |
| Backward-Outside | 37.5 | 36.7 | 36.8 | 39.4 | 75.9 | 78.2 | 29.5 | 31.8 |
| Forward | 5.5 | 4.7 | 1.5 | 2.5 | 3.0 | 3.5 | 1.6 | 1.9 |
| Forward-Inside | 14.9 | 16.9 | 3.3 | 6.4 | 5.3 | 8.3 | 2.1 | 2.6 |
| Forward-Outside | 5.6 | 3.9 | 1.8 | 1.3 | 4.6 | 5.7 | 3.0 | 3.3 |
| Inside | 36.7 | 37.7 | 22.3 | 18.3 | 46.4 | 44.9 | 13.1 | 12.0 |
| Outside | 24.1 | 21.3 | 15.6 | 16.5 | 35.5 | 39.2 | 14.3 | 13.9 |

3.3.2 - Subsequent game event percentages by deployment and zone. Here we have the likelihood of a game event occurring for each direction and can see how weak side deployments can generate events more often with backward-inside wins.

| Player | WDBE % | CF % | XGD | Clean % Among Wins | Win % Among Clean | Overall Win % | Strong-Side % |
|---|---|---|---|---|---|---|---|
| Andreas Athanasiou | 17.9% | 51.1% | 51.7% | 34.3% | 39.1% | 40.2% | 46.5% |
| Blake Wheeler | 19.2% | 53.2% | 58.3% | 40.1% | 44.2% | 44.7% | 51.8% |
| Nico Hischier | 19.4% | 53.5% | 54.8% | 49.5% | 44.6% | 42.4% | 53.2% |
| Jared McCann | 19.4% | 49.0% | 50.0% | 45.0% | 37.9% | 39.5% | 59.6% |
| Pavel Zacha | 20.7% | 49.2% | 51.3% | 36.6% | 41.1% | 44.6% | 51.2% |
| Logan Couture | 21.0% | 56.1% | 57.1% | 37.2% | 45.1% | 45.5% | 53.7% |
| Adrian Kempe | 21.5% | 50.9% | 53.1% | 38.7% | 35.6% | 38.8% | 61.2% |
| Alexander Wennberg | 21.7% | 57.1% | 57.4% | 43.6% | 45.9% | 46.9% | 49.5% |
| Joel Eriksson Ek | 21.8% | 43.6% | 46.2% | 40.9% | 40.0% | 41.1% | 58.7% |
| Jujhar Khaira | 21.9% | 46.6% | 43.0% | 45.8% | 49.7% | 47.2% | 59.8% |
| Connor McDavid | 22.4% | 55.9% | 60.2% | 45.7% | 42.2% | 43.3% | 54.4% |
| Jacob De La Rose | 22.5% | 44.3% | 39.3% | 40.8% | 45.9% | 44.1% | 61.8% |
| Derick Brassard | 22.8% | 55.8% | 59.0% | 46.9% | 48.2% | 48.0% | 50.6% |
| Alexander Kerfoot | 23.1% | 53.4% | 58.6% | 45.4% | 42.6% | 40.9% | 58.2% |
| Evgeni Malkin | 23.1% | 60.7% | 63.7% | 44.3% | 45.3% | 45.8% | 49.0% |
| Jack Eichel | 23.3% | 57.1% | 60.1% | 41.7% | 40.2% | 43.3% | 60.6% |
| Nick Schmaltz | 23.4% | 53.0% | 49.8% | 45.3% | 38.7% | 39.9% | 65.4% |
| Pierre-Luc Dubois | 23.4% | 58.8% | 57.0% | 42.9% | 42.1% | 43.3% | 50.3% |
| Mathew Barzal | 24.0% | 59.3% | 63.0% | 48.5% | 42.5% | 44.4% | 58.7% |
| Max Domi | 24.0% | 54.7% | 52.9% | 43.6% | 44.1% | 44.9% | 56.4% |

4 - Bottom 20 players from the deployment and win-type adjusted WDBE model (equation 2). Here we can observe how players with poor clean win percentages and differentials fail to generate value for their teams when they're taking the draw. These players may not be given optimal deployment opportunities, which means that strategy may need to change as well.